\documentclass[pra,aps,amsmath,amssymb,amsfonts,twocolumn,nofootinbib,floatfix]{revtex4}
\usepackage{}
\usepackage{amssymb}
\usepackage{bm,mathrsfs}
\usepackage{graphicx}
\usepackage{epsfig}
\usepackage{amsmath,bbm}
\usepackage{amsfonts,amssymb}
\usepackage{times}
\usepackage{verbatim}
\usepackage[sort&compress]{natbib}
\usepackage{amsmath}
\usepackage{bm}
\usepackage{float}
\usepackage{textgreek}
\usepackage{textcomp}
\allowdisplaybreaks[4]
\usepackage[colorlinks,breaklinks,linkcolor=blue,anchorcolor=blue,citecolor=blue,urlcolor=magenta]{hyperref}

\usepackage{color}
\definecolor{zzz}{rgb}{0.9,0.0,0.4}

\begin{document}

\title{Nonreciprocal Coupling Induced Non-Hermitian Entanglement in Optomechanical Systems}
\author{W. Y. Hu,$^{1}$ H. Yi,$^{1}$ H. N. Liu,$^{1}$ S. Y. Ge,$^{2}$ Cheng Shang,$^{3,}$\footnote{\textcolor{zzz}{Corresponding author: cheng.shang@riken.jp}} Yan-Hui Zhou,$^{4,} $\footnote{\textcolor{zzz}{Corresponding author: yanhuizhou@126.com }} and H. Z. Shen$^{1,}$\footnote{\textcolor{zzz}{Corresponding author: shenhz458@nenu.edu.cn }}}

\affiliation{$^1$Center for Quantum Sciences and School of Physics, Northeast Normal University, Changchun 130024, China\\
$^2$Laboratory of Quantum Information, University of Science and Technology of China, Hefei, Anhui 230026, China\\
$^3$Analytical Quantum Complexity RIKEN Hakubi Research Team,
RIKEN Center for Quantum Computing (RQC), 2-1 Hirosawa, Wako, Saitama 351-0198, Japan\\
$^4$Quantum Information Research Center and Jiangxi Province Key Laboratory of Applied Optical Technology, Shangrao Normal University, Shangrao 334001, China}

\begin{abstract}
Quantum entanglement, one of the most intrinsic properties of quantum mechanics, corresponds to a key resource for quantum information. Exceptional points (EPs) are special phase transition points in non-Hermitian systems. Nonreciprocity achieves one-way motion by breaking reciprocal symmetries. However, the connections among quantum entanglement, EPs, and nonreciprocity remain unexplored in non-Hermitian systems. In this paper, we investigate the manipulation of optomechanical entanglement by non-Hermitian coupling between two counter-propagating optical modes based on a whispering-gallery-mode (WGM) resonator coupled with two Rayleigh scatterers. The spinning resonator can induce opposite frequency shifts in two counter-propagating optical modes. Moreover, EPs can periodically appear by controlling the relative angle between two nanoparticles. The logarithmic negativity of entanglement reaches the maximum at EPs, leading to significant enhancement of optomechanical entanglement. In contrast, the entanglement is suppressed at non-EPs but can be recovered by the Sagnac-Fizeau shift. Specially, by driving the resonator at EPs while spinning it at the optimal rotation speed, we achieve switching between classical and quantum nonreciprocity.
\end{abstract}


\maketitle
\section{Introduction}
Quantum entanglement, the ``spooky'' interaction of quantum physics \cite{RHorodecki818652009,ESchrodinger315551935,AEinstein477771935}, characterizes nonclassical correlations and inseparability of nanoparticles regardless of their distance. It is a key resource in quantum information processing \cite{UAndersen43372010,FFlamini820160012018}, quantum communication \cite{RUrsin34812007,JWPan41010672001,DHan1723001032023}, quantum networks \cite{YRen1621005862022}, and quantum precision measurements \cite{HYu583432020,MLaHaye304742004,AMotazedifard180730402016}. Entanglement has been demonstrated in various physical systems including ions \cite{VKrutyanskiy1300508032023}, atomic ensembles \cite{XYLuo1290505032022,TvanLeent607692022}, photons \cite{XCYao62252012,ZGe10eadm75652024}, superconducting circuits \cite{PCampagneIbarcq1202005012018,PKurpiers5582642018,WNing1230605022019}, magnonic systems \cite{JLi1212036012018,BHussain1050637042022,MYu1242136042020,FeiWang1110337122025,ZZhang1110337172025,ZYFan1722008662023}, and cavity optomechanical (COM) devices \cite{DVitali980304052007,XMa1110335032025,YFJiao1251436052020,ZHHuang1100124232024,JHuang1060635062022,XWXu880223252013,PChen70131612025,TAPalomaki3427102013,RRiedinger5303132016,SMancini881204012002,LTian1102336022013,YDWang1102536012013,SBarzanjeh5704802019,DGLai1290636022022,RRiedinger5564732018,CFOckeloenKorppi5564782018,
SKotler3726222021,LMercierdeLepinay3726252021}. By utilizing efficient light-motion couplings \cite{MAspelmeyer8613912014}, COM systems have contributed to many applications, such as quantum transducers \cite{FLecocq1160436012016,ZShen1292436012022}, quantum memory \cite{MBKristensen1321008022024}, ground-state cooling \cite{JDTeufel4753592011,DGLai980238602018,YWang661242132023,DGLai1020115022020,JHuang1060135262022,YMLiu1130135232026}, weak-force sensing \cite{BBLi1027992021,EGavartin75092012,DMason157452019}, phonon lasing \cite{ISGrudinin1040839012010,GWang5732017,JZhang124792018,HJing1130536042014}, optomechanically induced transparency \cite{SWeis33015202010,AHSafaviNaeini472692011}, squeezed or entangled states \cite{YFJiao23011542024}, and normal-mode splitting \cite{Barzanjeh063850,Massimiliano073601,Dobrindt263602}.

Non-Hermitian systems \cite{ZhuyaoJin1130424522026,PXue1361700012026,ASergi910621082015,Agarwal012226,Barch094307} and EPs \cite{CMBender8052431998,CMBender709472007,VVKonotop880350022016,LFeng117522017,MAMiri363eaar77092019, CERuter61922010,HLu100140062018,Miao014403,Sticlet075133,Yang053712,HZShene715352026} have attracted intense interest in recent years. Two or more eigenstates coalesce at EPs, leading to a variety of unconventional effects observed in experiments, such as loss-induced coherence \cite{BPeng3463282014, AGuo1030939022009}, exotic topological states \cite{HYZhou35910092018}, photon blockade
induced by chirality and dissipation \cite{YLZuo1060437152022,RHuang1621004302022}, unidirectional lasing \cite{SLonghi5B12017}, unidirectional invisibility \cite{ZLin1062139012011}, resilient wireless energy transmission \cite{SAssawaworrarit5463872017}, laser mode selection \cite{LFeng3469722014,HHodaei3469752014}, coherent perfect absorption \cite{YSun1121439032014,CWang37312612021}, and skin effects \cite{Zhang045415,Kawabata021007,Okuma085428,Gliozz136503,Wang085418}. The EP effects in COM systems have been explored both theoretically and experimentally \cite{HXu537802016,HJing1130536042014,HLu80440202017,HJing733862017}, such as low-power phonon laser \cite{HJing1130536042014,HLu80440202017}, high-order EPs in COM systems \cite{HJing733862017}, and nonreciprocal COM devices \cite{HXu537802016}. Especially, by coupling the WGM resonator with two external nanoparticles, the periodic emergence of EPs was observed experimentally for tuning relative positions of nanoparticles \cite{BPeng11368452016}. The counterintuitive EP effects, such as modal chirality \cite{BPeng11368452016,QTCao1180339012017} and highly sensitive sensing \cite{WJChen5481922017,TTian1000438102019}, have been revealed in these devices.

Nonreciprocal devices achieve unidirectional light propagation by breaking Lorentz reciprocity of optical transmission \cite{Wang022216,Lei083604,Ramezani043901,Zhang744,Dong6193,Kim275,Xu044070,Manipatruni213903,
Shen657,Bernier604,Mercier034027,Fang465,Hafezi7672,Wang093901,Yang1021,Maayani569,Sounas774,Peng394,Yu013517,Jalas20137,Li013515,Yonglin043702}. An ideal WGM resonator possesses the high quality factor (Q-factor) and two degenerate cavity modes [clockwise (CW) and counterclockwise (CCW) modes] with same frequency. In the spinning resonator, the Sagnac effect \cite{POST475,Ruyong143901,Abhay1975} alters effective refractive index, inducing opposite frequency shifts for counter-propagating optical modes. Classical nonreciprocal transmission at specific frequencies has been experimentally demonstrated \cite{Maayani569}, while quantum nonreciprocal phenomena in spinning cavities have been theoretically proposed \cite{YFJiao1251436052020}. Based on mode frequency shifts in spinning cavities, nonreciprocal chaos \cite{DWZhang1040335222021} and photon blockade effects \cite{Li630,Jing033707,Chengdeng043723,Huang153601,Dong22343,Yu12847,Yunlan22020,I063828,Zhu053526,Shen013826,Xue4424}
can be realized.  

The influences of EPs on entanglement have been discussed in non-Hermitian systems \cite{YiBingQian1120435482025,ZigengLi31193822023,Wang013708,ZHYuan1100237322024,ZHYuan1110144332025}, while nonreciprocal entanglement \cite{YFJiao1251436052020} has been studied in COM devices. However, the connections among EPs, entanglement, and nonreciprocity remain unexplored in non-Hermitian COM systems. The above considerations lead us to investigate these issues: (i) Can optomechanical entanglement be manipulated via non-Hermitian coupling and enhanced at EPs? (ii) How does the Sagnac-Fizeau effect induced by the spinning resonator influence the entanglement at EPs and non-EPs? (iii) Can the system at EPs facilitate the switch between classical and quantum nonreciprocity through steering frequency tuning?

In this paper, we investigate entanglement at EPs in the non-Hermitian COM system coupled to a driven WGM resonator with two nanoparticles, where the coupling between CW and CCW traveling waves is nonreciprocal.  We find that EPs can periodically emerge in the system by adjusting relative positions of two nanoparticles. The entanglement can be enhanced as effective optomechanical coupling strength reaches the maximum at EPs. In contrast, the entanglement is suppressed at non-EPs but can be recovered by the Sagnac-Fizeau shift. In particular, the entanglement occurs in one direction while remaining uncorrelated in the other, giving rise to quantum nonreciprocity in the absence of classical nonreciprocity.

The paper is organized as follows. In Sec.~\ref{two}, the theoretical model and Hamiltonian are described for WGM resonator coupled with  mechanical mode. In Sec.~\ref{three}, we derive motion equations and linearized effective non-Hermitian Hamiltonian. In Sec.~\ref{four}, we study the behavior of entanglement at EPs through varying system parameters. In Sec.~\ref{five}, we study the WGM resonator spinning at the rotation speed along its axis and investigate its impact on entanglement. The main conclusions are summarized in Sec.~\ref{seven}.

\section{MODEL and Hamiltonian}\label{two}

\begin{figure}
\centerline{
\includegraphics[height=3.5cm,width=8.5cm]{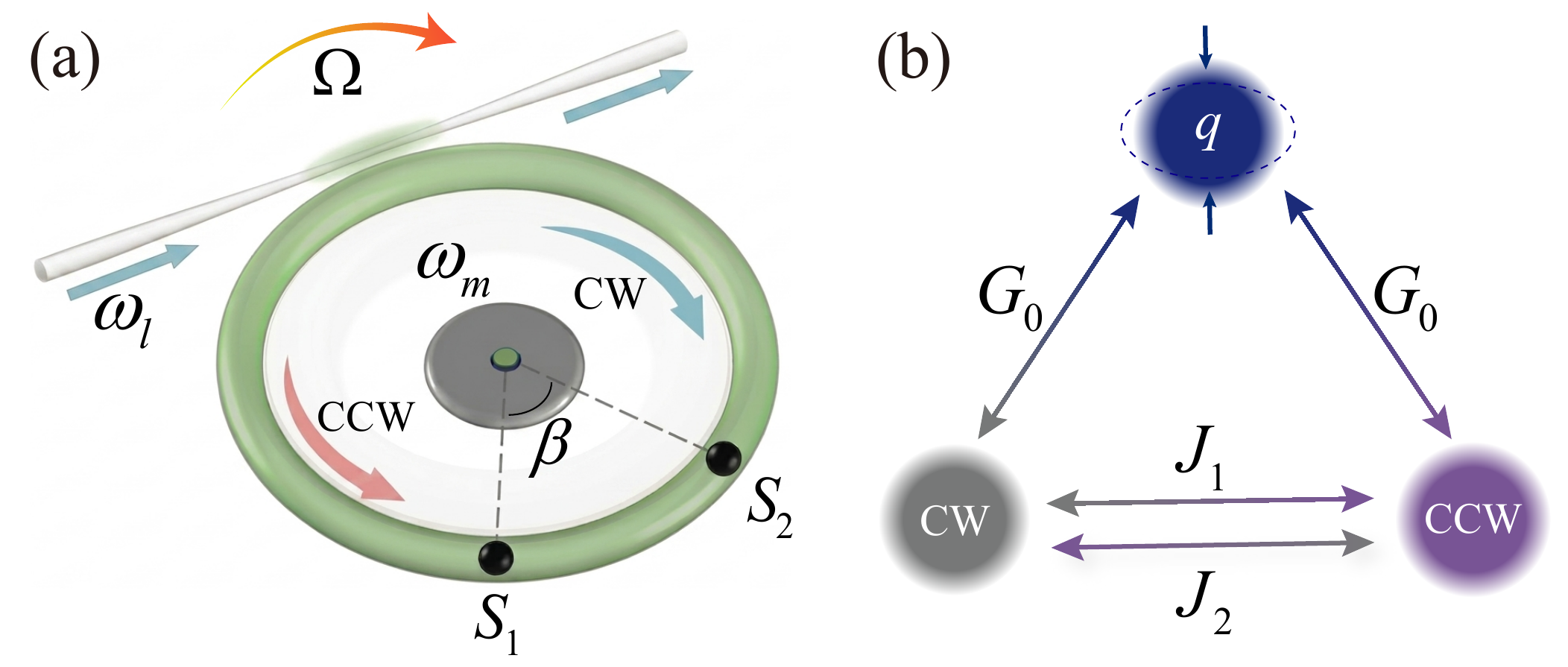}}
\caption{The model diagram of optomechanical entanglement. (a) Optomechanics in a resonator is perturbed by two nanoparticles in the WGM field, where red and blue arrows represent CCW and CW modes, respectively. The resonator driven by laser with frequency $\omega_l$ supports a mechanical mode at frequency $\omega_m$. $\beta$ is the relative angle between two nanoparticles denoted by $S_1$ and $S_2$. $\Omega$ is angular velocity of resonator. (b) Nonreciprocal couplings (${J_1}$, ${J_2}$) between optical CW and CCW modes. The optomechanical couplings ($G_0$) between optical (CW, CCW) and mechanical modes ($q$).}
\label{a1.eps}
\end{figure}

In Fig.~\ref{a1.eps}, we consider a WGM resonator consisting of both CW and CCW optical modes with mechanical mode. Two silica nanotips as Rayleigh scatterers are placed in the evanescent field of resonator, which are fabricated by the wet etching of tapered fiber tips prepared by heating and stretching standard optical fibers. The position of each nanoparticle is controlled by nanopositioner, which tunes relative position and effective size of nanoparticle in WGM fields. The non-Hermitian coupling strength between CCW and CW optical modes induced by two nanoparticles can be characterized by scattering rates, where ${J_1} = {\epsilon_1} + {\epsilon_2}{e^{ - i2m\beta }}$ and ${J_2} = {\epsilon_1}+ {\epsilon_2}{e^{i2m\beta }}$correspond to coupling coefficients for photon conversion from CCW (CW) mode to CW (CCW) mode, respectively. $\epsilon_i$ $(i=1,2)$ is complex frequency splitting induced by $i$th scatterer, which depends on the volume of $i$th nanoparticle within WGM fields and is tuned by controlling distance between nanoparticle and resonator with nanopositioners. The integer $m$ denotes azimuthal mode number, while steering angle $\beta$ can bring the system to EPs, as already observed experimentally \cite{RHuang1621004302022}.
Considering the case of coherent driving laser input along CCW direction, the total Hamiltonian of system reads
$\hat{H}' ={{\hbar\omega_{\circlearrowleft}}{\hat{a}^{\dag}_{\circlearrowleft}}{\hat{a}_{\circlearrowleft}}}
+{\hbar\omega_{\circlearrowright}}{\hat{a}^{\dag}_{\circlearrowright}}{\hat{a}_{\circlearrowright}}
+ \hbar \omega_m ( \hat{p}^2 + \hat{q}^2 ) / 2
+{\hbar}{J_{1}}{\hat{a}^{\dag}_{\circlearrowright}}{\hat{a}_{\circlearrowleft}}
+{\hbar}{J_{2}}{\hat{a}^{\dag}_{\circlearrowleft}}{\hat{a}_{\circlearrowright}}
{-\hbar{G}_{0}}{\hat{a}^{\dag}_{\circlearrowleft}}{\hat{a}_{\circlearrowleft}}{\hat{q}}
-{\hbar{G}_{0}}{\hat{a}^{\dag}_{\circlearrowright}}{\hat{a}_{\circlearrowright}}{\hat{q}}
+{i\hbar\varepsilon({\hat{a}^{\dag}_{\circlearrowleft}{e}^{-i\omega_{l}t}}-{\hat{a}_{\circlearrowleft}}{e}^{i\omega_{l}t})}$,
where ${\hat a_j^\dag} ({\hat a_j}) (j = \circlearrowleft, \circlearrowright)$ represents the creation (annihilation) operator for cavity mode. Moreover, the nanoparticles also introduce the self-scattering frequency shift for each cavity mode, represented by \(\hbar(\epsilon_1 + \epsilon_2)\hat{a}_j^\dagger \hat{a}_j\). $\hat q (\hat p)$ is dimensionless mechanical displacement (momentum) operator. $G_0 = (\omega_c / R) \times \sqrt{\hbar / m_{eff} \omega_m}$ denotes single-photon COM coupling rate \cite{MAspelmeyer8613912014} with mass $m_{eff}$ and radius $R$ of resonator. $\omega_c$ is resonance frequency for stationary resonator. The field amplitude of driving laser is $|\varepsilon| = \sqrt{2\kappa P / \hbar\omega_l}$, where $P$ and $\kappa$ are input laser power and optical decay rate, respectively. When driving laser is input in the CCW direction, the optical field frequencies of CCW and CW modes are ${\omega _\circlearrowleft} = {\omega _c} + {{\rm{\Delta }}_F}$ and ${\omega _\circlearrowright} = {\omega _c} - {{\rm{\Delta }}_F}$, where the Sagnac-Fizeau shift $\Delta_F = \Omega (n R \omega_c / c) [1 - 1/n^2 - (\lambda/n) (dn/d\lambda)]$ \cite{Maayani569,Malykin1229,Jiang064037,Shi28114}. $n$ denotes refractive index of medium. $c(\lambda)$ indicates the speed of light in vacuum (at wavelength $\lambda$). The dispersion term $dn/d\lambda$ is small in typical materials \cite{Maayani569,Malykin1229}. The Hamiltonian $\hat{H}'$ in the rotating frame becomes
\begin{equation}
\begin{aligned}
\hat H =&{\hbar\Delta^{\prime}_{\circlearrowleft}}{\hat{a}^{\dag}_{\circlearrowleft}}{\hat{a}_{\circlearrowleft}}
+{\hbar\Delta^{\prime}_{\circlearrowright}}{\hat{a}^{\dag}_{\circlearrowright}}{\hat{a}_{\circlearrowright}}
+\frac{\hbar {\omega _m}}{2}( {{{\hat p}^2} + {{\hat q}^2}} )\\
&+{\hbar}{J_{1}}{\hat{a}^{\dag}_{\circlearrowright}}{\hat{a}_{\circlearrowleft}}
+{\hbar}{J_{2}}{\hat{a}^{\dag}_{\circlearrowleft}}{\hat{a}_{\circlearrowright}}
{-\hbar{G}_{0}}{\hat{a}^{\dag}_{\circlearrowleft}}{\hat{a}_{\circlearrowleft}}{\hat{q}}
-{\hbar{G}_{0}}{\hat{a}^{\dag}_{\circlearrowright}}{\hat{a}_{\circlearrowright}}{\hat{q}}\\
&+{i\hbar\varepsilon({\hat{a}^{\dag}_{\circlearrowleft}}-{\hat{a}_{\circlearrowleft}})},
\label{2}
\end{aligned}
\end{equation}
where $\Delta '_j = {{\rm{\Delta }}_j} + {\rm{Re}}( {{\epsilon_1} + {\epsilon_2}} )$. The Hamiltonian \(\hat{H}\) in Eq.~(\ref{2}) describes the coherent evolution of system, thereby incorporating only the real part of complex frequency splitting \(\mathrm{Re}(\epsilon_1 + \epsilon_2)\). The imaginary part \(\mathrm{Im}(\epsilon_1 + \epsilon_2)\) representing the dissipative scattering losses induced by nanoparticles is excluded at the Hamiltonian stage and later incorporated into Heisenberg-Langevin equations via the total optical dissipation rate. The frequency detunings of rotating and stationary cavity optical modes relative to driving field are defined as $\Delta_j = \omega_j - \omega_l$ and $\Delta = \omega_c - \omega_l$, respectively. The nanoparticles induce frequency splitting of optical modes with corresponding eigenfrequencies obtained from Hamiltonian describing optical modes and their interaction \cite{Lu095401} ${\omega_{\pm}}={{\omega_{c}}-i\kappa_{c}+\epsilon_{1}+\epsilon_{2}\pm}\sqrt {\epsilon^{2}_{1}+\epsilon^{2}_{2}+2\epsilon_{1}\epsilon_{2}\cos(2m\beta)}$. Since $\epsilon_1$ and $\epsilon_2$ are complex numbers, the complex eigenvalue splitting $\Delta\omega = \omega_{+} - \omega_{-}$ is allowed to be zero by tuning the angle $\beta$, which brings the system to the EP. For $J_1 \neq J_2$ and $J_1J_2 = 0$, the system has only one eigenvalue and one eigenvector, indicating the emergence of EP. In this case, when $\Delta\omega=0$, we derive $|\epsilon_1| = |\epsilon_2|$ and obtain the critical value of $\beta$ as $\beta_{c} = l\pi / 2m \mp [\arg(\epsilon_1) - \arg(\epsilon_2)] / 2m$  $(l = \pm 1, \pm 3, \ldots)$, where ``$+$'' is selected for $J_2=0$, while ``$-$'' is taken for $J_1=0$. By tuning $\beta$ \cite{BPeng11368452016, WJChen5481922017}, $\beta_c$ emerges periodically.
\begin{figure}
\centerline{
\includegraphics[height=4cm,width=8.5cm]{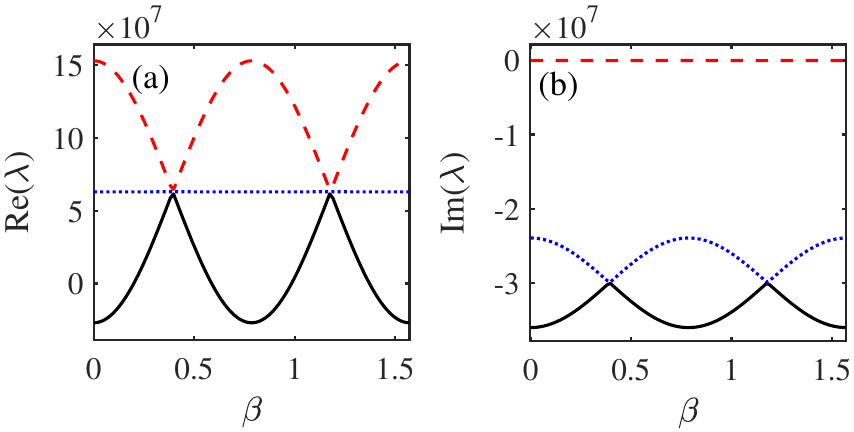}}
\caption{The complex eigenfrequency $\lambda$ of Hamiltonian as a function of phase angle $\beta$. The real and imaginary parts of eigenvalues for system are shown in (a) and (b), respectively. The parameters chosen are $\omega_m = 63~\mathrm{MHz}$, $\kappa_c = 30\,\text{MHz}$, $\gamma_m = 5.2~\mathrm{kHz}$, ${\epsilon_1}/{\kappa _c} = 1.5 - 0.1i$, ${\epsilon_2}/{\kappa _c} = 1.4999 - 0.101489i$, and $m=4$.}
\label{2heng.eps}
\end{figure}

\section{Dynamical Evolution of System}\label{three}

When the dissipation and input noises are considered \cite{Gardiner2004}, Heisenberg-Langevin equations with Eq.~(\ref{2}) give
\begin{equation}
\begin{aligned}
\dot{\hat{a}}_{\circlearrowleft}=&-(i\Delta_{\circlearrowleft}+\kappa)\hat{a}_{\circlearrowleft}
-iJ_{2}\hat{a}_{\circlearrowright}+iG_{0}\hat{a}_{\circlearrowleft}\hat{q}+\varepsilon
+\sqrt {2\kappa}{\hat{a}^{in}}_{\circlearrowleft},\\
\dot{\hat{a}}_{\circlearrowright}=&-(i\Delta_{\circlearrowright}+\kappa)\hat{a}_{\circlearrowright}
-iJ_{1}\hat{a}_{\circlearrowleft}+iG_{0}\hat{a}_{\circlearrowright}\hat{q}
+\sqrt {2\kappa}{\hat{a}^{in}}_{\circlearrowright},\\
\dot{\hat{q}}=&\omega_{m}\hat{p},\\
\dot{\hat{p}}=&-\omega_{m}\hat{q}+G_{0}\sum \limits_{j = \circlearrowleft,\circlearrowright} \hat a_j^\dag {\hat a_j}-\gamma_{m}\hat{p}+\hat{\xi},
\label{6}
\end{aligned}
\end{equation}
where the total optical dissipation is given by $\kappa  = {\kappa _c} - {\rm{Im}}( {{\epsilon_1} + {\epsilon_2}} )$. $\kappa_c$ and $\gamma_m$ denote optical and mechanical damping rates, respectively. ${\hat{a}^{in}}_{j}$ denotes the input noise operator, which is assumed to have a zero mean value and satisfies $\langle{\hat{a}^{in}}_{j}(t){\hat{a}^{in\dagger}}_{j}(t^{\prime})\rangle=\delta(t-t^{\prime})$ and $\langle{\hat{a}^{in\dagger}}_{j}(t){\hat{a}^{in}}_{j}(t^{\prime})\rangle=0$. Moreover, $\hat{\xi}$ denotes Brownian noise operator for mechanical mode meeting $\langle\hat{\xi}(t) \hat{\xi}(t^{\prime})\rangle = (\gamma_{m}/\omega_{m}) \int (d \omega/2 \pi) e^{-i \omega(t-t^{\prime})} [\coth (\hbar \omega / 2 k_{B} T)+1]$, where $k_{B}$ and $T$ represent the Boltzmann constant and bath temperature, respectively. Under the constraints of high quality factor for mechanical mode ($Q_{m} = \omega_{m} / \gamma_{m} \gg 1$), $\hat{\xi}(t)$ becomes delta correlated \cite{DVitali980304052007}
$\langle\hat{\xi}^{\mathrm{in}}(t) \hat{\xi}^{\mathrm{in}}(t^{\prime}) + \hat{\xi}^{\mathrm{in}}(t^{\prime}) \hat{\xi}^{\mathrm{in}}(t)\rangle / 2 = \gamma_{m}(2 n_m+1) \delta(t-t^{\prime})$, where ${n_m}={[{\rm{exp}}( {\hbar {\omega _m}/{k_B}T} ) - 1]^{ - 1}}$ is thermal phonon number.
We linearize Eq.~(\ref{6}) by expressing each operator as a sum of steady-state values and quantum fluctuation operators ${\hat a_j} = {\alpha _j} + \delta {\hat a_j}$, $\hat q = {q_s} + \delta \hat q$, and $\hat p = {p_s} + \delta \hat p$ with
\begin{equation}
\begin{aligned}
0=&-[ {i( {\rm \Delta '_\circlearrowleft} - {G_0}{q_s}} ) + \kappa  ]{\alpha _\circlearrowleft}{ - i{J_2}{\alpha _\circlearrowright} + \varepsilon},\\
0=&{-[ {i( {{\rm\Delta '_\circlearrowright} - {G_0}{q_s}} ) + \kappa } ]{\alpha _\circlearrowright}}{ - i{J_1}{\alpha _\circlearrowleft}},\\
0=& {\omega _m}{p_s},\\
0=&-{{\omega _m}{q_s} - {\gamma _m}{p_s}}{ + {G_0}( {| {{\alpha _\circlearrowright}{|^2} + } |{\alpha _\circlearrowleft}{|^2}} )}.
\label{10}
\end{aligned}
\end{equation}
The validity of linearization relies on strong driving conditions, which enables the analysis of quantum fluctuations around steady state. By defining effective detuning ${{{\rm{\tilde \Delta '}}}_j} = \Delta '_j  - {G_0}{q_s}$, we obtain steady-state solutions
$\alpha _\circlearrowleft = [( i{\rm{\tilde \Delta }}^{\prime}_ \circlearrowright+ \kappa )\varepsilon ] / [( i{\rm{\tilde \Delta }}^{\prime}_\circlearrowright + \kappa )( i{\rm{\tilde \Delta }}^{\prime}_ \circlearrowleft+ \kappa ) + J_1 J_2 ]$,
$\alpha _\circlearrowright = ( -i J_1 \alpha _\circlearrowleft) / ( i{\rm{\tilde \Delta }}^{\prime}_\circlearrowright + \kappa )$,
$q_s = ( G_0 / \omega_m ) ( |\alpha _\circlearrowright|^2 + |\alpha _\circlearrowleft|^2 )$, and $p_s = 0$, which lead to effective detuning as ${\rm{\tilde \Delta '}}_j = \Delta '_j - ({G_0}^2/{\omega _m}) ( |\alpha _\circlearrowright|^2 + |\alpha _\circlearrowleft|^2 )$. Neglecting second-order fluctuation terms, the linearized quantum Langevin equations for fluctuation operators can be obtained as
\begin{equation}
\begin{aligned}
\delta{\dot{\hat{a}}}_{\circlearrowleft}=&-( {i{{{\rm{\tilde \Delta }}}^{\prime}}_\circlearrowleft + \kappa } )\delta {\hat a_\circlearrowleft} - i{J_2}\delta {\hat a_\circlearrowright}+i{G_0}{\alpha _\circlearrowleft}\delta \hat q + \sqrt {2\kappa } \hat a_\circlearrowleft^{{\rm{in}}},\\
\delta{\dot{\hat{a}}}_{\circlearrowright}=&-( {i{{{\rm{\tilde \Delta }}}^{\prime}}_ \circlearrowright+ \kappa } )\delta {\hat a_\circlearrowright} - i{J_1}\delta {\hat a_\circlearrowleft} + i{G_0}{\alpha _\circlearrowright}\delta \hat q + \sqrt {2\kappa } \hat a_\circlearrowright^{{\rm{in}}},\\
\delta{\dot{\hat{q}}}=&{\omega _m}\delta \hat p,\\
\delta{\dot{\hat{p}}}=& - {\omega _m}\delta \hat q - {\gamma _m}\delta \hat p \\
&+ {G_0}( {\alpha _\circlearrowright^*\delta {{\hat a}_\circlearrowright} + \alpha _\circlearrowleft^*\delta {{\hat a}_\circlearrowleft} + {\alpha _\circlearrowright}\delta \hat a_\circlearrowright^\dag  + {\alpha _\circlearrowleft}\delta \hat a_\circlearrowleft^\dag } ) + \hat\xi.
\label{12}
\end{aligned}
\end{equation}
We define linearized coupling strengths ${G_j} = \sqrt 2 {G_0}{\alpha _j}$. Based on Eq.~\eqref{12}, the linearized effective Hamiltonian of three-mode COM system under the rotating-wave approximation (RWA) can be written as
\begin{equation}
\begin{aligned}
{\hat H_{{\rm{eff}}}} = \left( {\begin{array}{*{20}{c}}
{{{\tilde \Delta }_\circlearrowleft ^\prime - i\kappa }}&{{J_2}}&{ - \frac{{{G_\circlearrowleft}}}{2}}\\
{{J_1}}&{{{\tilde \Delta }_\circlearrowright ^\prime - i\kappa }}&{ - \frac{{{G_\circlearrowright}}}{2}}\\
{ - \frac{{G_\circlearrowleft^*}}{2}}&{ - \frac{{G_\circlearrowright^*}}{2}}&{{\omega _m} - i\frac{{{\gamma _m}}}{2}}
\end{array}} \right),
\label{13}
\end{aligned}
\end{equation}
whose derivation can be found in Appendix~\ref{B}. For such $3\times3$ matrix, solving the eigenvalues is complex, so we employ numerical methods to solve eigenvalues $\lambda$ of Hamiltonian matrix Eq.~(\ref{13}). In Fig.~\ref{2heng.eps}, we find that EPs appear at $ \beta\approx0.4$ and $\beta\approx1.2$, where both real and imaginary parts of eigenvalues $\lambda$ of non-Hermitian Hamiltonian coalesce.
By defining quadrature fluctuation operators vector ${u}( t ) = ( {\delta {{\hat X}_\circlearrowleft},\delta {{\hat Y}_\circlearrowleft},\delta {{\hat X}_\circlearrowright},\delta {{\hat Y}_\circlearrowright},\delta \hat q,\delta \hat p} )^\mathrm{T}$ and noise operators vector ${z}( t ) = ( {\sqrt {2\kappa } \hat X_\circlearrowleft^{{\rm{in}}},\sqrt {2\kappa } \hat Y_\circlearrowleft^{{\rm{in}}},\sqrt {2\kappa } \hat X_\circlearrowright^{{\rm{in}}},\sqrt {2\kappa } \hat Y_\circlearrowright^{{\rm{in}}},0,\hat \xi } )^\mathrm{T}$ with components $\delta {\hat X_j} = (1/\sqrt{2})( \delta {\hat a_j} + \delta \hat a_j^\dag )$,
$\delta {\hat Y_j} = (i/\sqrt{2})( \delta \hat a_j^\dag  - \delta {\hat a_j} )$,
$\hat X_j^{\rm{in}} = (1/\sqrt{2})( \hat a_j^{\rm{in}} + \hat a_j^{\rm{in}\dag} )$, and $\hat Y_j^{\rm{in}} = (i/\sqrt{2})( \hat a_j^{\rm{in}\dag} - \hat a_j^{\rm{in}} )$.
Equation~\eqref{12} can be written as $\dot u( t ) = Au( t ) + z( t )$, where drift matrix $A$ is given by
\begin{equation}
\resizebox{0.95\hsize}{!}{$
\begin{aligned}
A =
\left( {\begin{array}{*{20}{c}}
{ - \kappa }&{{\rm{\tilde \Delta }}_\circlearrowleft^{\prime}}&{{\rm{Im}}( {{J_2}} )}&{{\rm{Re}}( {{J_2}} )}&{ - G_\circlearrowleft^y}&0\\
{ - {\rm{\tilde \Delta }}_\circlearrowleft^{\prime}}&{ - \kappa }&{ - {\rm{Re}}( {{J_2}} )}&{{\rm{Im}}( {{J_2}} )}&{G_\circlearrowleft^x}&0\\
{{\rm{Im}}( {{J_1}} )}&{{\rm{Re}}( {{J_1}} )}&{ - \kappa }&{{\rm{\tilde \Delta }}_\circlearrowright^{\prime}}&{ - G_\circlearrowright^y}&0\\
{ - {\rm{Re}}( {{J_1}} )}&{{\rm{Im}}( {{J_1}} )}&{ - {\rm{\tilde \Delta }}_\circlearrowright^{\prime}}&{ - \kappa }&{G_\circlearrowright^x}&0\\
0&0&0&0&0&{{\omega _m}}\\
{G_\circlearrowleft^x}&{G_\circlearrowleft^y}&{G_\circlearrowright^x}&{G_\circlearrowright^y}&{ - {\omega _m}}&{ - {\gamma _m}}
\end{array}} \right)
\end{aligned}
$},
\label{199}
\end{equation}
where ${\rm{\tilde \Delta '}}= \Delta '_j - {G_0}{q_s}$ is effective optical detuning. $G^{x}_{j}$ and $G^{y}_{j}$ represent real and imaginary parts of effective COM coupling rate (${G_j} = G_j^x + iG_j^y$), respectively. According to Routh-Hurwitz criterion \cite{DeJesus5288}, the system is stable and reaches steady state only when all eigenvalues of $A$ in Eq.~(\ref{199}) have negative real parts.
Due to Gaussian properties of quantum noise and linearity of quantum Langevin equations (see the validity of linearization approximation near EPs in Appendix~\ref{A}), the quantum state of system is characterized by $6 \times 6$ covariance matrix $V$ with matrix elements ${V_{kl}} = \langle{{u_k}( t ){u_l}( t ) + {u_l}( t ){u_k}( t )}\rangle / 2$.
The steady-state covariance matrix $V$ is determined by Lyapunov equation $AV + V{A^\mathrm{T}} = -D$ \cite{DVitali980304052007}, where $D = {\rm{Diag}}[ {\kappa ,\kappa ,\kappa ,\kappa ,0,{\gamma _m}( {2{n_m} + 1} )} ]$ represents diffusion matrix with matrix elements $D_{ij}$ defined by $\langle \hat{z}_i(t)\hat{z}_j(t') + \hat{z}_j(t')\hat{z}_i(t) \rangle/2 = D_{ij} \delta(t-t')$. To quantify bipartite entanglement between optical and mechanical modes, the logarithmic negativity $E_N$ \cite{GVidal650323142002,MBPlenio950905032005,RSimon8427262000} is defined as entanglement measure
\begin{equation}
\begin{aligned}
{E_N} = \max [0, - \ln (2{\xi ^ - })],
\label{34}
\end{aligned}
\end{equation}
where $\xi^{-} = 2^{-1/2} \{ \mathcal{Q} - [ \mathcal{Q}^2 - 4 \det({V_{sub}}) ]^{1/2} \}^{1/2}$ with $\mathcal{Q} = \det(\phi_1) + \det(\phi_2) - 2 \det(\phi_3)$. $\phi_1$, $\phi_2$, and $\phi_3$ are $2 \times 2$ submatrices derived from covariance matrix ${V_{sub}}$. ${V_{sub}} = \left( \begin{smallmatrix} \phi_1 & \phi_3 \\ \phi_3^\mathrm{T} & \phi_2 \end{smallmatrix} \right)$ is the $2 \times 2$ block form.
\begin{figure}
\centerline{
\includegraphics[height=3.6cm,width=8.7cm]{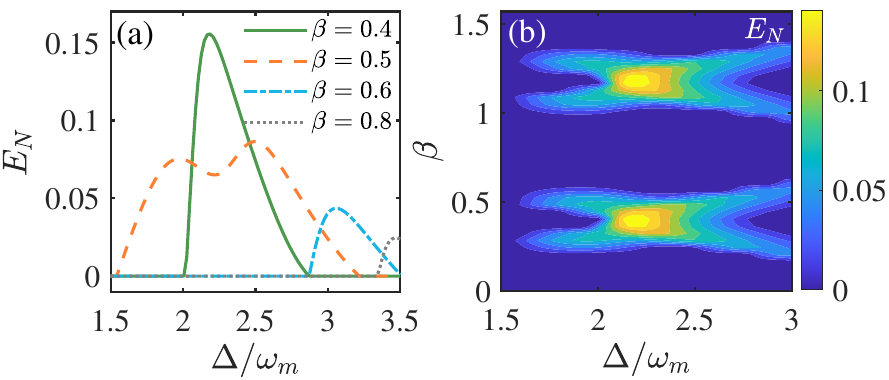}}
\caption{The logarithmic negativity in Eq.~\eqref{34} for (a,b) optomechanical entanglement is plotted as a function of detuning $\Delta$ and phase angle $\beta$. The green solid, orange dashed, blue dash-dotted, and grey dotted lines respectively correspond to $\beta = 0.4$, $\beta = 0.5$, $\beta = 0.6$, and $\beta = 0.8$. The parameters chosen are $\omega_m = 63~\mathrm{MHz}$, $\omega_c = 1.22~\mathrm{PHz}$, $\kappa_c = 30\,\text{MHz}$, $\gamma_m = 5.2~\mathrm{kHz}$, ${\epsilon_1}/{\kappa _c} = 1.5 - 0.1i$, ${\epsilon_2}/{\kappa _c} = 1.4999 - 0.101489i$, $m=4$, $m_{eff} = 10~\mathrm{ng}$, $R = 1.1~\mathrm{mm}$, $\lambda = 1.55~\mu\mathrm{m}$, $P=20 $mW, and $T = 130~\mathrm{mK}$.}
\label{3xin.eps}
\end{figure}

\section{Discussions on EPs mediated non-Hermitian Optomechanical Entanglement}\label{four}
The driving laser is modulated by mechanical mode and scattered into Stokes and anti-Stokes sidebands. By tuning cavity mode to resonate with driving laser sidebands, corresponding scattering process can be enhanced, generating optomechanical entanglement \cite{TAPalomaki3427102013, RRiedinger5303132016}. We choose to input red-detuned laser, where beam-splitter interaction dominates in COM system. According to amplitude distribution of optical modes in the resonator, the Stokes process is suppressed while the anti-Stokes process can be enhanced. The anti-Stokes sideband resonates with cavity mode, thereby generating optomechanical entanglement. Specifically, the suppression of Stokes process is determined by the choice of red-detuned driving laser. In this regime, the frequency of input laser \( \omega_l \) is lower than cavity resonance frequency \( \omega_c \) with detuning \( \Delta = \omega_c - \omega_l \) matching mechanical frequency \( \omega_m \). In the anti-Stokes process, a laser photon absorbs the energy of a mechanical phonon, up-converting scattered photon to the higher-frequency sideband at \( \omega_l + \omega_m \), which matches the cavity resonance, leading to enhanced the anti-Stokes process. In the Stokes process, a laser photon releases energy to create a mechanical phonon, resulting in down-converted scattered photon at the lower-frequency sideband \( \omega_l - \omega_m \) (corresponding to $\omega_c-2\omega_m$), which is shifted further away from cavity resonance frequency $\omega_c$. The Stokes process is suppressed due to optical modes being mainly distributed around the resonance frequency.
\begin{figure}
\centerline{
\includegraphics[height=4.2cm,width=4.2cm]{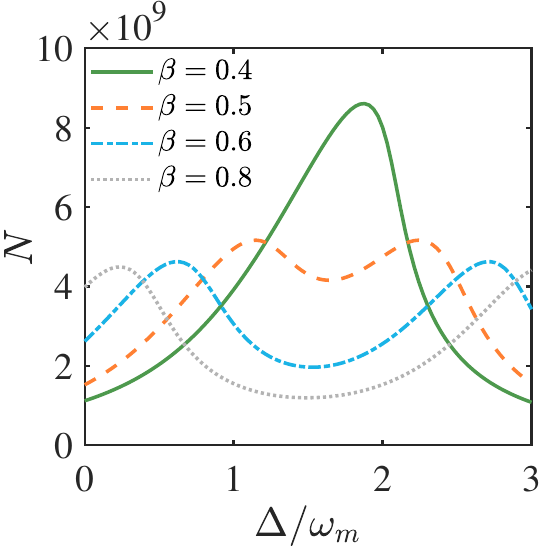}}
\caption{The mean intracavity photon number $N$ as a function of detuning $\Delta$ and phase angle $\beta$. The green solid, orange dashed, blue dash-dotted, and grey dotted lines respectively correspond to $\beta = 0.4$, $\beta = 0.5$, $\beta = 0.6$, and $\beta = 0.8$. The parameters are chosen to be the same as those in Fig.~\ref{3xin.eps}.
}
\label{4xin.eps}
\end{figure}

We first consider the manipulation of entanglement via phase angle $\beta$ in the stationary cavity $(\Delta_F = 0)$. Figure~\ref{3xin.eps} shows logarithmic negativity $E_N$ as a function of detuning $\Delta$ for different phase angles $\beta$ between two nanoparticles. In Fig.~\ref{3xin.eps}(a), the green solid line demonstrates that at the EP $(\beta\approx0.4)$, $E_N$ reaches the maximum $(E_N\approx0.16)$ at $\Delta/{\omega _m} \approx 2.18$, indicating the strongest entanglement. At this point, two optical modes share common eigenfrequency. As phase angle $\beta$ moves away from EP $(\beta=0.5, 0.6,$ and $0.8)$, the eigenfrequencies of two optical modes split, which suppresses the entanglement. The entanglement vanishes at $\beta=0.8$, where frequency splitting reaches the maximum. In Fig.~\ref{3xin.eps}(b), we find that $E_N$  exhibits periodic variations with changing $\beta$. Specifically, $E_N$ vanishes far from EPs and enhances when approaching EPs. $E_N$ reaches  the maximum at EPs $\beta\approx0.4$ and $\beta\approx1.2$. The physical mechanism behind this phenomenon can be connected to the eigenvalue spectrum in Fig.~\ref{2heng.eps}. The Hamiltonian of system is written in the rotating frame, where the corresponding anti-Stokes sideband frequency is \(\omega_m\), while the Stokes sideband corresponds to \(-\omega_m\). In Fig.~\ref{2heng.eps}(a), the real parts of hybridized optical mode eigenvalues are degenerate at the anti-Stokes sideband frequency at EPs, which is marked by blue dotted line. The precise frequency matching ensures resonance between the anti-Stokes sideband and cavity mode, thereby maximizing the entanglement. Away from the EP, the eigenvalues split further apart, moving away from the anti-Stokes sideband, resulting in weaker entanglement.
\begin{figure}
\centerline{
\includegraphics[height=3.6cm,width=8.5cm]{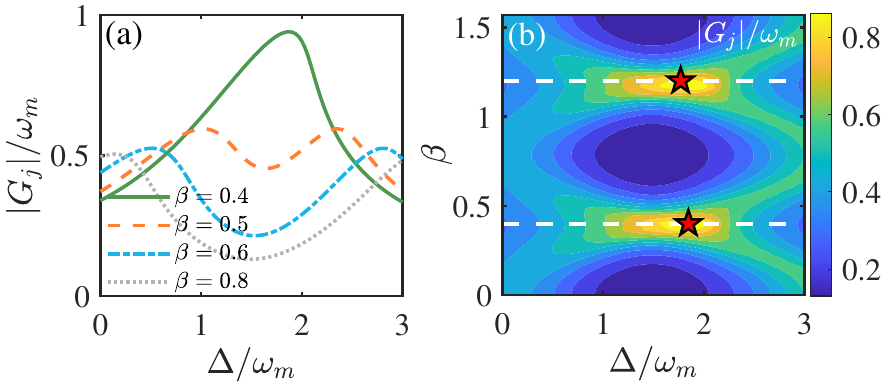}}
\caption{The effective optomechanical coupling strength $| {{G_j}} |/{\omega _m}$ as a function of phase angle $\beta$ and detuning ${\rm{\Delta }}/{\omega _m}$. The green solid, orange dashed, blue dash-dotted, and grey dotted lines respectively correspond to $\beta = 0.4$, $\beta = 0.5$, $\beta = 0.6$, and $\beta = 0.8$. The parameters are chosen to be the same as those in Fig.~\ref{3xin.eps}.
}
\label{5xin.eps}
\end{figure}

Figures~\ref{4xin.eps} and \ref{5xin.eps} respectively show mean intracavity photon number $N$ and effective optomechanical coupling strength $|G_j|/\omega_m$ as a function of phase angle $\beta$ and detuning $\Delta$. In Fig.~\ref{4xin.eps}, mean intracavity photon number $N$ reaches the maximum at ${\rm{\Delta }}/{\omega _m} \approx 1.8$ for EP ($\beta\approx0.4$). In Fig.~\ref{5xin.eps}, optical modes coalesce for EPs ($\beta\approx0.4$ and $\beta\approx1.2$), where effective optomechanical coupling strength reaches the maximum $| {{G_j}} |/{\omega _m}\approx0.94$ at ${\rm{\Delta }}/{\omega _m}\approx1.8$. At non-EPs, frequency splitting occurs between CW and CCW modes. Backscattering alters the intracavity optical field distribution and coupling between optical modes, leading to reduced mean intracavity photon number, which decreases the effective optomechanical coupling strength and suppresses the entanglement. It is worth noting that there is a discrepancy between the detuning position for the maximum mean intracavity photon number (i.e., maximum optomechanical coupling strength at \(\Delta/\omega_m \approx 1.8\) in Figs.~\ref{4xin.eps} and \ref{5xin.eps}) and the detuning position for the optimal entanglement at \(\Delta/\omega_m \approx 2.18\) in Fig.~\ref{3xin.eps}. This phenomenon can be explained by differences between the classical steady-state solutions and the dynamics of quantum fluctuations. Specifically, the detuning position for the maximum mean intracavity photon number $N_j \equiv |\alpha_j|^2$ (i.e., maximum optomechanical coupling strength ${G_j} = \sqrt 2 {G_0}{\alpha _j}$) is determined by steady-state equations corresponding to Eq.~(\ref{10}), which includes the final term $G_0( |\alpha_\circlearrowright|^2 + |\alpha_\circlearrowleft|^2 )$ corresponding to the nonlinear optomechanical coupling terms (i.e., $-\hbar G_{0} \hat{a}^{\dag}_{\circlearrowleft} \hat{a}_{\circlearrowleft} \hat{q} - \hbar G_{0} \hat{a}^{\dag}_{\circlearrowright} \hat{a}_{\circlearrowright} \hat{q}$) from the Hamiltonian in Eq.~(\ref{2}). Solving for the steady-state mean intracavity photon number yields a transcendental equation due to the nonlinearity of Eq.~(\ref{10}), which determines the detuning position for the maximum mean intracavity photon number (i.e., maximum optomechanical coupling strength). Moreover, the entanglement is calculated based on the time-dependent linear quantum ﬂuctuation equation~(\ref{12}), where the second-order fluctuation terms are neglected to eliminate the nonlinear terms and make Eq.~(\ref{12}) linear. The drift matrix \( A \) in Eq.~(\ref{199}) is derived from time-dependent linear quantum ﬂuctuation equation~(\ref{12}) to compute covariance matrix \( V \) and logarithmic negativity \( E_N \) in Eq.~(\ref{34}), which determines the detuning position for optimal entanglement. The discrepancy exists between the optimal detuning position for the maximum mean intracavity photon number (i.e., maximum optomechanical coupling strength) and that for the optimal entanglement due to these different properties.

\section{non-Hermitian Quantum Entanglement in Spinning COM Systems}\label{five}
\begin{figure}
\centerline{
\includegraphics[height=3.8cm,width=8.8cm]{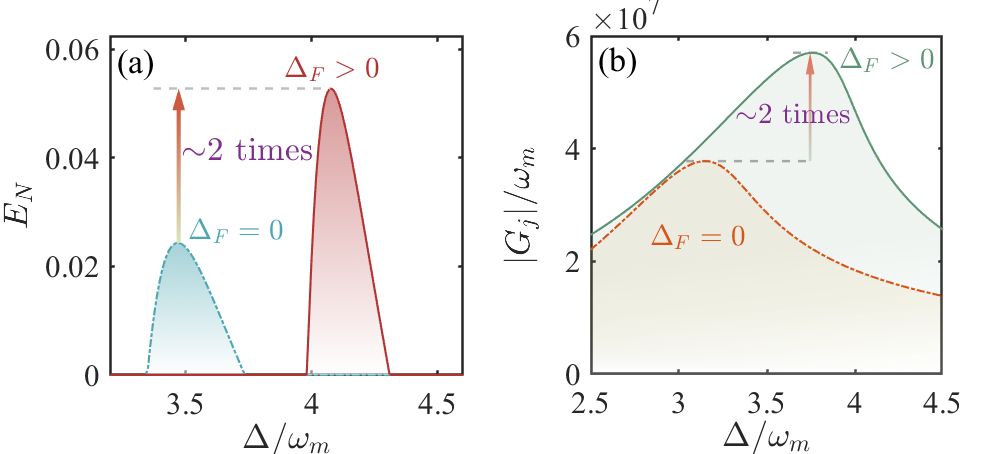}}
\caption{The recovery of optomechanical entanglement away from EPs in spinning resonator. The blue dash-dotted and red solid lines correspond to stationary cavity with \(\Delta_F = 0\) and spinning cavity with \(\Delta_F > 0\) for (a), respectively. The orange dash-dotted and green solid lines correspond to stationary cavity with \(\Delta_F = 0\) and spinning cavity with \(\Delta_F > 0\) for (b), respectively. The parameters chosen are $n = 1.48$, $\beta=0.8$, $\Omega=23\, \text{kHz}$, $\omega_m = 63~\mathrm{MHz}$, $\omega_c = 1.22~\mathrm{PHz}$, $\kappa_c = 30\,\text{MHz}$, $\gamma_m = 5.2~\mathrm{kHz}$, ${\epsilon_1}/{\kappa _c} = 1.5 - 0.1i$, ${\epsilon_2}/{\kappa _c} = 1.4999 - 0.101489i$, $m=4$, $m_{eff} = 10~\mathrm{ng}$, $R = 1.1~\mathrm{mm}$, $\lambda = 1.55~\mu\mathrm{m}$, $P=20 $mW, and $T = 130~\mathrm{mK}$.}
\label{fig6abxin.pdf}
\end{figure}

We consider the WGM resonator spinning along its axis at the angular velocity with ${{\rm{\Delta }}_F} \ne 0$. We find that rotation can suppress quantum decoherence away from EPs. We rotate the WGM resonator with $\beta=0.8$ at the rotation speed $\Omega=23 \,\text{kHz}$ in the CCW direction $({{\rm{\Delta }}_F} > 0)$ in Fig.~\ref{fig6abxin.pdf}. In Fig.~\ref{fig6abxin.pdf}(a), the peak value of $E_N$ increases to twice that of stationary cavity $({{\rm{\Delta }}_F} = 0)$, indicating the recovery of entanglement. In Fig.~\ref{fig6abxin.pdf}(b), the peak effective optomechanical coupling strength $| {{G_j}} |$ of spinning cavity also increases to twice that of stationary cavity. These phenomena arise from resonance frequency splitting between CW and CCW modes induced by random-defect-induced backscattering, which suppresses the entanglement. However, the Sagnac-Fizeau shift $\Delta_F$ induced by CCW rotation of resonator compensates for mismatch between the anti-Stokes sideband and cavity mode caused by frequency splitting at non-EPs. The compensation enhances anti-Stokes scattering process and recovers optomechanical coupling strength and entanglement.

\begin{figure}
\centerline{
\includegraphics[height=3.65cm,width=4.5cm]{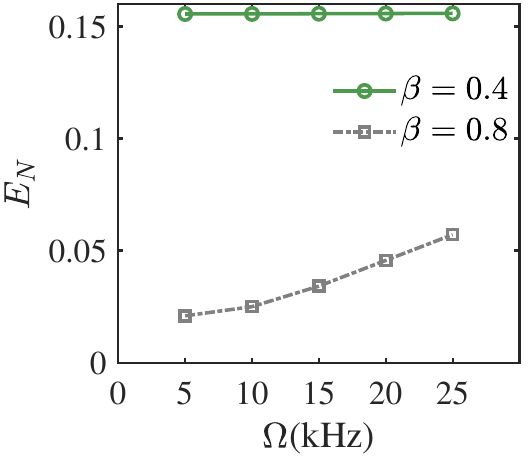}}
\caption{The logarithmic negativity $E_N$ as a function of rotation speed $\Omega$. The cases of $\beta = 0.4$ and $\beta = 0.8$ are represented by green circle solid line and gray square dash-dotted line, respectively. The parameters chosen are $\beta=0.4$ and $\beta = 0.8$, while other parameters are the same as in Fig.~\ref{fig6abxin.pdf}.
}
\label{7yuan.eps}
\end{figure}

We investigate influences of rotation speed $\Omega$ for entanglement, comparing conditions at EPs and non-EPs regimes. Figure~\ref{7yuan.eps} shows the logarithmic negativity \( E_N \) as a function of rotation speed \( \Omega \) at phase angles \( \beta = 0.4 \) and \( \beta = 0.8 \). Away from EPs (gray square dash-dotted line), the entanglement is sensitive to rotation speed \( \Omega \). The entanglement \( E_N \approx 0.02 \) at \( \Omega = 5 \, \text{kHz} \), while when \( \Omega \) increases to \( 25 \, \text{kHz} \), \( E_N \) becomes three times larger. The underlying mechanism is that at the non-EP (\( \beta = 0.8 \)), the backscattering from two nanoparticles causes large frequency splitting between CW and CCW modes, which suppresses the resonance condition between the anti-Stokes sideband and cavity mode. When the resonator spins, the introduced Sagnac-Fizeau shift compensates for the frequency mismatch caused by backscattering. Therefore, the compensation effect improves as rotation speed approaches optimal value, leading to the recovery of entanglement. In contrast, the entanglement remains almost unchanged as rotation speed increases at the EP ($\beta = 0.4$). The flat green circle solid line indicates theoretical immunity to Sagnac-Fizeau effect at the EP. The underlying mechanism is that at the EP (\( \beta = 0.4 \)), the system is at a special transition point where the eigenvalues of CW and CCW modes coalesce. Specially, the coalesced optical mode at the EP can exhibit chirality \cite{ZTang131672024,ZHLi1100135152024}. The Sagnac-Fizeau effect shifts resonance frequency without inducing new mode splitting due to chirality.

It is important to emphasize that high rotation speeds (e.g., \(\Omega = 23\)-\(25 \, \text{kHz}\)) in Figs.~\ref{fig6abxin.pdf} and \ref{7yuan.eps} are employed to illustrate the asymptotic behavior of Sagnac-Fizeau effect. Spinning the microresonator at such extreme speeds translating to 1.5 million RPM presents challenges in the experiment. The centrifugal stress would alter the optical and mechanical properties of the WGM resonator. Moreover, aerodynamic drag and mechanical vibrations would disrupt the phase stability, making it difficult to maintain the precise sub-wavelength spatial overlap with stationary external silica nanotips required for the EPs at \(\beta \approx 0.4\) and \(\beta \approx 1.2\). Therefore, our high speed rotation represents a theoretical result. To elevate our model to current state-of-the-art standards, the nonreciprocal mechanism developed in this paper can be experimentally realized by incorporating symmetry-breaking mechanisms that do not require physical rotation. Recent studies demonstrate that the Barnett frequency shift induces nonreciprocal quantum correlations by reversing the polarity of induced magnetization~\cite{TXLu1110137132025}. Moreover, utilizing purely synthetic gauge fields generates tunable nonreciprocal entanglement without moving parts~\cite{DGLai1290636022022}. Finally, to address thermal decoherence issue, transitioning to ultra-high-frequency molecular vibrations (\(\sim\)30THz) would allow entanglement to persist at much higher temperatures (300 K to 1000 K)~\cite{EKBerinyuy7134173132025,EKBerinyuy1408062025}.

\begin{figure}
\centerline{
\includegraphics[height=4cm,width=8.5cm]{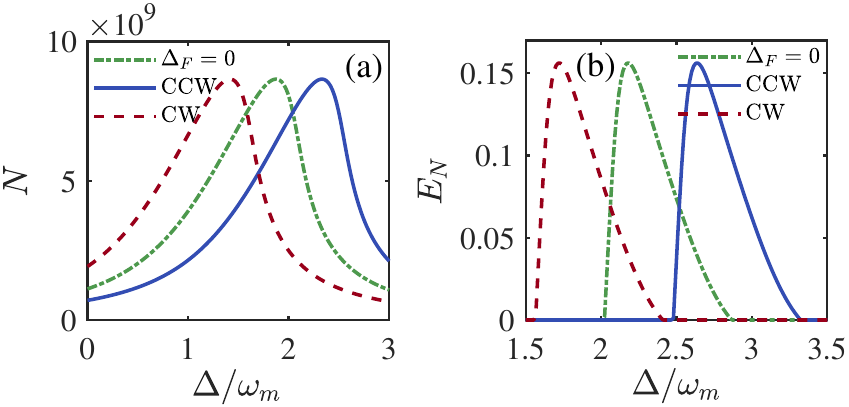}}
\caption{Sagnac effect at EPs in spinning cavity. (a) mean intracavity photon number $N$ and (b) logarithmic negativity $E_N$ as a function of detuning $\Delta/\omega_{m}$ for different driving laser input directions. The green dash-dotted, blue solid, and red dashed lines correspond to ${{\rm{\Delta }}_F} = 0$, CCW $({{\rm{\Delta }}_F} > 0)$, and CW $({{\rm{\Delta }}_F} < 0)$, respectively. The parameters chosen are $\beta=0.4$ and $\Omega=8 \,\text{kHz}$, while other parameters are the same as in Fig.~\ref{fig6abxin.pdf}.}
\label{8yuan.eps}
\end{figure}

We investigate the Sagnac effect at EPs in spinning COM systems. In Fig.~\ref{8yuan.eps}, The spectral shifts in the peaks of mean intracavity photon number $N$ and entanglement $E_N$ originate from cavity mode red shift induced by effective optomechanical coupling. The green dash-dotted lines represent ${{\rm{\Delta }}_F} = 0$, where the input direction of driving laser shows no influences on the entanglement. Due to frequency shift induced by nanoparticles, $E_N$ reaches the maximum at ${\rm{\Delta }}/{\omega _m} \approx 2.18$. For spinning resonator, the entanglement depends on drive direction. When the cavity is driven from either the CW or CCW direction, the frequency detuning corresponding to the optimal entanglement condition exhibits symmetric shifts.
\begin{figure}
\centerline{
\includegraphics[height=11.5cm,width=5.5cm]{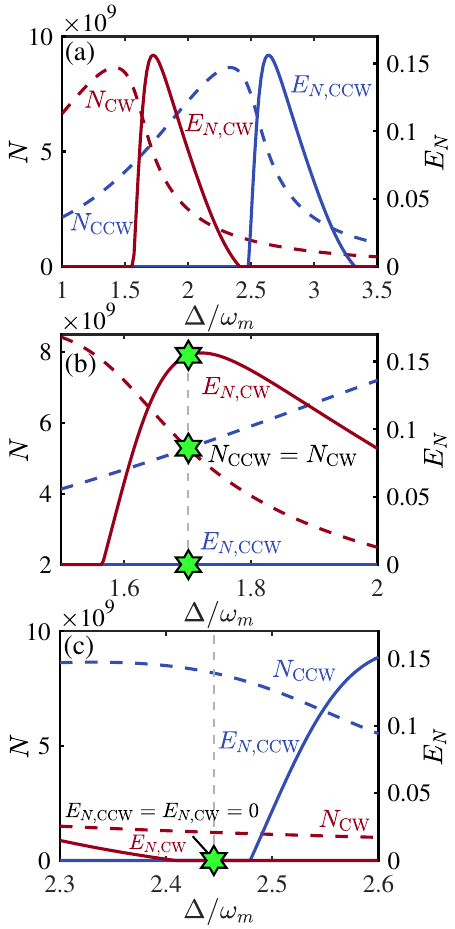}}
\caption{Classical and quantum nonreciprocity of spinning cavity at EP. The mean intracavity photon number $N$ (red and blue dashed lines) and logarithmic negativity $E_N$ (red and blue solid lines) as functions of detuning $\Delta/\omega_{m}$ for different driving laser input directions. The parameters chosen are $\beta=0.4$ and $\Omega=8 \,\text{kHz}$, while other parameters are the same as in Fig.~\ref{fig6abxin.pdf}.}
\label{fig9abcxin.pdf}
\end{figure}

Figure~\ref{fig9abcxin.pdf}(a) illustrates the overall trends of mean intracavity photon number $N$ (red and blue dashed lines) and the entanglement $E_N$ (red and blue solid lines) as a function of detuning $\Delta / \omega_m$. It provides a necessary reference for choosing specific detuning regimes, while Figs.~\ref{fig9abcxin.pdf}(b) and \ref{fig9abcxin.pdf}(c) respectively focus on $\Delta / \omega_m \approx 1.7$ and $\Delta / \omega_m \approx 2.45$ to show the switching between classical and quantum nonreciprocity. In Fig.~\ref{fig9abcxin.pdf}(b), we observe that quantum nonreciprocity can be achieved in absence of classical nonreciprocity. Specifically, there is no classical nonreciprocity (i.e., $N_{\mathrm{CCW}} = N_{\mathrm{CW}}$) at ${\rm{\Delta }}/{\omega _m} \approx 1.7$, but quantum nonreciprocity exists (i.e., $E_{N,\mathrm{CCW}} = 0$ and $E_{N,\mathrm{CW}} \neq 0$). Figure~\ref{fig9abcxin.pdf}(c) demonstrates that classical nonreciprocity can be observed in absence of quantum nonreciprocity. The logarithmic negativity of CW and CCW modes satisfies $E_{N,\mathrm{CCW}} = E_{N,\mathrm{CW}} = 0$ at ${\rm{\Delta }}/{\omega _m} \approx 2.45$, indicating the absence of quantum nonreciprocity. However, the classical nonreciprocity exists as the mean intracavity photon number of CCW mode exceeds that of CW mode  $N_{\mathrm{CCW}} \gg N_{\mathrm{CW}}$. 
The physical mechanism behind the classical and quantum nonreciprocity stems from two differences. The classical nonreciprocity (characterized by massive asymmetry in the mean intracavity photon number, e.g., $N_{\mathrm{CCW}} \gg N_{\mathrm{CW}}$) is dependent on the nonlinear optomechanical interaction and steady-state solutions corresponding to Eq.~(\ref{10}). Moreover, the quantum nonreciprocity (e.g., $E_{N,\mathrm{CCW}}=0$ and $E_{N,\mathrm{CW}}\neq0$) is determined by time-dependent linear quantum ﬂuctuation equation~(\ref{12}). The massive asymmetry in the classical steady-state photon populations does not directly map to asymmetry in the quantum fluctuations responsible for entanglement due to these different properties. This discovery suggests the possibility of switching between quantum and classical nonreciprocity in optical nonreciprocal devices through frequency tuning.

\begin{figure}
\centerline{
\includegraphics[height=4cm,width=4.8cm]{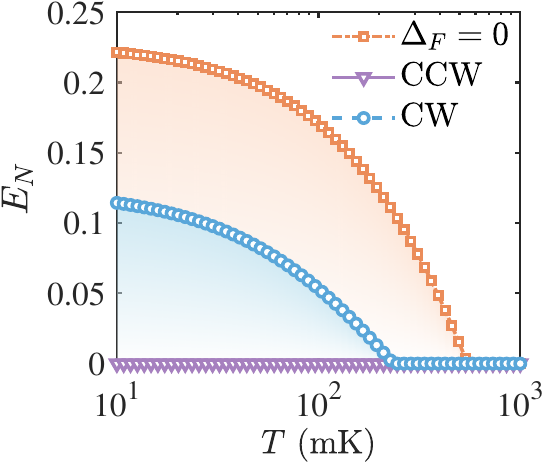}}
       \caption{The logarithmic negativity $E_N$ as a function of temperature $T$. The orange square, purple triangle, and blue circle correspond to ${{\rm{\Delta }}_F} = 0$, CCW $({{\rm{\Delta }}_F} > 0)$, and CW $({{\rm{\Delta }}_F} < 0)$, respectively. The parameters chosen are $\beta=0.4$, $\Delta/\omega_m = 2.18$, and $\Omega=8 \,\text{kHz}$, while other parameters are the same as in Fig.~\ref{fig6abxin.pdf}.} \label{fig10xin.pdf}
\end{figure}

Temperature is an important factor for entanglement in open quantum systems due to the rapid decoherence and destruction of quantum entanglement caused by thermal noise. In Fig.~\ref{fig10xin.pdf}, we plot logarithmic negativity $E_N$ as a function of temperature $T$ ranging from $10$~mK to $1000$~mK. The detuning is fixed at $\Delta/\omega_m = 2.18$, which corresponds to the entanglement peak of stationary cavity ($\Delta_F = 0$). We find that $E_N$ for both the stationary cavity (orange square) and the spinning cavity driven in the CW direction (blue circle) decreases with temperature $T$ due to thermal decoherence. As $T$ rises to $1000$~mK, the entanglement for $\Delta_F = 0$ can exist at temperatures exceeding $500$~mK before disappearing. The entanglement for the spinning cavity driven in the CW (${{\rm{\Delta }}_F} < 0$) direction can persist up to $200$~mK. Specifically, Fig.~\ref{fig10xin.pdf} shows that the system exhibits quantum nonreciprocity in the low-temperature regime. The CCW $({{\rm{\Delta }}_F} > 0)$ mode (purple triangle) remains unentangled ($E_N = 0$) at $\Delta/\omega_m = 2.18$ regardless of $T$. The entanglement can exist in the CW direction while being prohibited in the opposite CCW direction.

\section{Conclusion}\label{seven}
In summary, we have studied non-Hermitian quantum entanglement in the spinning COM system coupled with two nanotips, which supports CW and CCW optical modes with a mechanical radial breathing mode. The two optical modes exhibit non-Hermitian coupling mediated by two nanoparticles, while optical and mechanical modes interact via radiation pressure. We find that the entanglement can be enhanced at EPs. Away from EPs, the entanglement is suppressed but can be recovered by Sagnac-Fizeau shift. By utilizing the Sagnac effect induced by the rotating cavity in opposite directions, we can achieve the switching between classical and quantum nonreciprocity.
These results indicate the benefits of using non-Hermitian coupling and enhance our understanding of entanglement in COM systems. The scheme developed might be extended to systems with common reservoir \cite{SYGuan81022022,JYang210440562024}, Rabi model \cite{YHChen1330336032024,YHChen752024}, and molecular induced transparency \cite{BYin1110435072025}.

\section*{AUTHOR CONTRIBUTIONS}
This project was conceived by C. Shang and H. Z. Shen. The methodology was developed by W. Y. Hu, H. Yi, H. N. Liu, S. Y. Ge,  C. Shang, Yan-Hui Zhou, and H. Z. Shen. The project was supervised by C. Shang, Yan-Hui Zhou, and H. Z. Shen. The original draft of the manuscript was written by W. Y. Hu, S. Y. Ge, and H. Z. Shen. The manuscript was reviewed and edited by W. Y. Hu, H. Yi, H. N. Liu, S. Y. Ge,  C. Shang, Yan-Hui Zhou, and H. Z. Shen.

\section*{ACKNOWLEDGMENTS}
We thank the referees for their constructive comments that helped improve our work. This work was supported by Science and Technology Development Plan Project of Jilin Province (Grant No.~20250102007JC), National Natural Science Foundation of China (NSFC) under Grants No.~12274064 and No. 12374333, and the Hakubi Projects of RIKEN.

\section*{CONFLICTS OF INTEREST}
The authors declare no conflicts of interest.

\section*{DATA AVAILABILITY STATEMENT}
The data that support the findings of this study are available from the corresponding author upon reasonable request.

\section*{CODE AVAILABILITY STATEMENT}
Code used to generate data in this study is available from the corresponding author upon reasonable request.\\

\noindent \textbf{Correspondence} and requests for materials should be addressed to C. Shang or H. Z. Shen.

\appendix


\section{Derivation of linearized effective Hamiltonian in Eq.~(\ref{13})}\label{B}

To derive the effective Hamiltonian $\hat{H}_{\text{eff}}$ in Eq.~(\ref{13}), we begin with linearized equations for fluctuation operators given in Eq.~(\ref{12}). Rewriting $\delta\hat{q}$ and $\delta\hat{p}$ in terms of $\delta\hat{q} = (\delta\hat{b} + \delta\hat{b}^\dagger)/\sqrt{2}$ and $\delta\hat{p} = i(\delta\hat{b}^\dagger - \delta\hat{b})/\sqrt{2}$, Eq.~(\ref{12}) becomes
\begin{equation}
\resizebox{1\hsize}{!}{$
\begin{aligned}
\delta\dot{\hat{a}}_\circlearrowleft &= -(i\tilde{\Delta}'_\circlearrowleft + \kappa)\delta\hat{a}_\circlearrowleft - iJ_2\delta\hat{a}_\circlearrowright + i\frac{G_\circlearrowleft}{2}(\delta\hat{b} + \delta\hat{b}^\dagger) + \sqrt {2\kappa } \hat a_\circlearrowleft^{{\rm{in}}}, \\
\delta\dot{\hat{a}}_\circlearrowright &= -(i\tilde{\Delta}'_\circlearrowright + \kappa)\delta\hat{a}_\circlearrowright - iJ_1\delta\hat{a}_\circlearrowleft + i\frac{G_\circlearrowright}{2}(\delta\hat{b} + \delta\hat{b}^\dagger) + \sqrt {2\kappa } \hat a_\circlearrowright^{{\rm{in}}}, \\
\delta\dot{\hat{b}} &= \frac{1}{\sqrt{2}}[ \omega_m\delta\hat{p} - i\omega_m\delta\hat{q} - i\gamma_m\delta\hat{p} + iG_0 \sum_{j=\circlearrowleft,\circlearrowright} (\alpha_j^*\delta\hat{a}_j + \alpha_j\delta\hat{a}_j^\dagger)] + \hat\xi.
\end{aligned}
$}
\label{B1}
\end{equation}
Under the near resonance condition ($\tilde{\Delta}_j^{\prime} \approx \omega_m$) and weak coupling ($|G_j| \ll \omega_m$), we apply the RWA to neglect the rapidly oscillating terms in Eq.~(\ref{B1}) and get
\begin{equation}
\begin{aligned}
\delta\dot{\hat{a}}_{\circlearrowleft} &= -(i\tilde{\Delta}_{\circlearrowleft}^{\prime} + \kappa)\delta\hat{a}_{\circlearrowleft} - iJ_2\delta\hat{a}_{\circlearrowright} + i\frac{G_{\circlearrowleft}}{2}\delta\hat{b} + \sqrt {2\kappa } \hat a_\circlearrowleft^{{\rm{in}}}, \\
\delta\dot{\hat{a}}_{\circlearrowright} &= -(i\tilde{\Delta}_{\circlearrowright}^{\prime} + \kappa)\delta\hat{a}_{\circlearrowright} - iJ_1\delta\hat{a}_{\circlearrowleft} + i\frac{G_{\circlearrowright}}{2}\delta\hat{b} + \sqrt {2\kappa } \hat a_\circlearrowright^{{\rm{in}}}, \\
\delta\dot{\hat{b}} &= -\left(i\omega_m + \frac{\gamma_m}{2}\right)\delta\hat{b} + i\frac{G_{\circlearrowleft}^*}{2}\delta\hat{a}_{\circlearrowleft} + i\frac{G_{\circlearrowright}^*}{2}\delta\hat{a}_{\circlearrowright} + \hat\xi.
\label{B2}
\end{aligned}
\end{equation}
By substituting Eq.~(\ref{B2}) into the matrix form of linearized quantum Langevin equations $\dot{h}(t) = -i\hat{H}_{\text{eff}} h(t) + w(t)$ with vector $h(t) = (\delta\hat{a}_{\circlearrowleft}, \delta\hat{a}_{\circlearrowright}, \delta\hat{b})^{\text{T}}$ and $w(t) = (\sqrt{2\kappa}\hat{a}_{\circlearrowleft}^{\text{in}}, \sqrt{2\kappa}\hat{a}_{\circlearrowright}^{\text{in}}, \hat{\xi})^{\text{T}}$, we obtain Eq.~(\ref{13}).

\section{The Analysis of Petermann Factor and the Validity of Linearization Approximation near EPs}
\label{A}

The quantum fluctuation analysis relies on linearizing the Heisenberg-Langevin equations around steady state, neglecting second-order fluctuation terms. At the EP, eigenvalues and eigenvectors coalesce while the Petermann factor diverges, leading to massive amplification of excess noise. If the system is tuned to the exact EPs ($\beta_{c_1} \approx 0.39285976125377924939$ and $\beta_{c_2} \approx 1.1782579246553255281$ satisfying $\beta_{c} = l\pi / 2m \mp [\arg(\epsilon_1) - \arg(\epsilon_2)] / 2m$ and taking parameters in Fig.~\ref{2heng.eps}), the massive excess noise will break the condition $|\delta\hat{a}| \ll |\alpha|$ required for first-order linearization.
However, we show that our parameters $\beta = 0.4$ and $\beta = 1.2$ are not exact EPs, but rather values chosen near EPs. The Petermann factor remains finite due to the slight parameter offset from exact EPs. Below, we provide the analysis of Petermann factor and prove the validity of first-order linearization near EPs. We consider the right and left eigenvectors of $\hat{H}_{\text{eff}}$ in Eq.~(\ref{13}) defined by $\hat{H}_{\text{eff}} |R_v\rangle = \lambda_v |R_v\rangle$ and $\hat{H}_{\text{eff}}^\dagger |L_v\rangle = \lambda_v^* |L_v\rangle$, where $\lambda_v$ is the complex eigenvalue. For each mode $v$, the Petermann factor is defined as
\begin{equation}
K_v = \frac{\langle L_v | L_v \rangle \langle R_v | R_v \rangle}{|\langle L_v | R_v \rangle|^2}.
\label{L2} 
\end{equation}
In contrast, $K_v > 1$ signals nonorthogonality, while $K_v \to \infty$ indicates the EP, consistent with the self-orthogonality of coalescing eigenvectors \cite{Ghosh2026013516,Berry200363,Heiss2012444016}. Notably, the divergence of Petermann factor reflecting self-orthogonality of coalescing eigenvectors is  associated with enhanced noise sensitivity in non-Hermitian systems.
The eigenvalue $\lambda_+$ is determined by characteristic equation $\det(\hat{H}_{\text{eff}} - \lambda_+ I) = 0$. The corresponding unnormalized right eigenvector $|R_+\rangle$ and left eigenvector $\langle L_+|$ can be derived as
\begin{equation}
\begin{aligned}
|R_+\rangle &= \begin{pmatrix} 
G_\circlearrowleft(\tilde{\Delta}_\circlearrowright^\prime - i\kappa - \lambda_+) - G_\circlearrowright J_2 \\ 
G_\circlearrowright(\tilde{\Delta}_\circlearrowleft^\prime - i\kappa - \lambda_+) - G_\circlearrowleft J_1 \\ 
2[(\tilde{\Delta}_\circlearrowleft^\prime - i\kappa - \lambda_+)(\tilde{\Delta}_\circlearrowright^\prime - i\kappa - \lambda_+) - J_1 J_2] 
\end{pmatrix}, \\[2ex]
\langle L_+| &= \begin{pmatrix} 
G_\circlearrowleft^*(\tilde{\Delta}_\circlearrowright^\prime - i\kappa - \lambda_+) - G_\circlearrowright^* J_1 \\ 
G_\circlearrowright^*(\tilde{\Delta}_\circlearrowleft^\prime - i\kappa - \lambda_+) - G_\circlearrowleft^* J_2 \\ 
2[(\tilde{\Delta}_\circlearrowleft^\prime - i\kappa - \lambda_+)(\tilde{\Delta}_\circlearrowright^\prime - i\kappa - \lambda_+) - J_1 J_2] 
\end{pmatrix}^{\mathrm{T}}.
\label{L3} 
\end{aligned}
\end{equation}
By substituting Eq.~(\ref{L3}) into Eq.~(\ref{L2}), we find $K_{\beta_{c_1}} \approx K_{\beta_{c_2}} \approx 9.5 \times 10^4$ for the exact EPs occurring at $\beta_{c_1} \approx 0.39285976125377924939$ and $\beta_{c_2} \approx 1.1782579246553255281$ taking parameters in Fig.~\ref{2heng.eps}. As the number of decimal places in the values of $\beta_{c_1}$ and $\beta_{c_2}$ continues to increase, $K_{\beta_{c_1}}$ and $K_{\beta_{c_2}}$ can continue to increase until $K_{\beta_{c_1}} \approx K_{\beta_{c_2}} \to \infty$. Using the parameters $\beta = 0.4$ and $\beta = 1.2$ that we actually used when plotting, we obtain $K_{\beta=0.4} \approx 1.0003$ and $K_{\beta=1.2} \approx 1.000004$, which proves that at our chosen points, the eigenvectors are orthogonal ($K_v \approx 1$), protecting the system from massive amplification of excess noise.

To validate the first-order linearization, we rely on the fundamental condition in cavity optomechanics \cite{MAspelmeyer8613912014}, which requires the steady-state quantum fluctuations to be negligibly small compared to the steady-state amplitude for the CW mode, i.e., $\langle \delta\hat{a}_{\circlearrowright}^\dagger \delta\hat{a}_{\circlearrowright} \rangle_{\text{s}} \ll |\alpha_{\circlearrowright}|^2$. Therefore, we define the dimensionless parameter
\begin{equation}
\eta = \frac{\sqrt{\langle \delta\hat{a}_{\circlearrowright}^\dagger \delta\hat{a}_{\circlearrowright} \rangle_{\text{s}}}}{|\alpha_{\circlearrowright}|}.
\label{11}
\end{equation}
The linearized quantum Langevin equation for the cavity fluctuation operator $\delta\hat{a}_{\circlearrowright}$ in Eq.~(\ref{12}) is given by
\begin{equation}
\delta\dot{\hat{a}}_{\circlearrowright} = -(i\tilde{\Delta}'_{\circlearrowright} + \kappa)\delta\hat{a}_{\circlearrowright} - i J_1 \delta\hat{a}_{\circlearrowleft} + i \frac{G_{\circlearrowright}}{\sqrt{2}} \delta\hat{q} + \sqrt{2\kappa}\hat{a}_{\circlearrowright}^{\text{in}}.
\end{equation}
Applying adiabatic elimination to the fast optical mode ($\kappa \gg \gamma_m$) by setting $\delta\dot{\hat{a}}_{\circlearrowright} \to 0$ and utilizing $J_1=0$ at the EP, we obtain
\begin{equation}
\delta\hat{a}_{\circlearrowright} = \frac{i (G_{\circlearrowright}/\sqrt{2})\delta\hat{q} + \sqrt{2\kappa}\hat{a}_{\circlearrowright}^{\text{in}}}{i\tilde{\Delta}'_{\circlearrowright} + \kappa}.
\end{equation}
Defining the mechanical position operator as $\hat{q} = (\hat{b} + \hat{b}^\dagger)/\sqrt{2}$, its steady-state variance is given by $\langle \delta \hat{q}^2 \rangle_{\text{s}} = n_m + 1/2$. Due to $n_m \gg 1$ at $T = 130\text{ mK}$, we get $\langle \delta\hat{q}^2 \rangle_{\text{s}} \approx n_m$. The steady-state photon fluctuation can be written as
\begin{equation}
\langle \delta\hat{a}_{\circlearrowright}^\dagger \delta\hat{a}_{\circlearrowright} \rangle_{\text{s}} \approx \frac{1}{2} \frac{|G_{\circlearrowright}|^2}{\tilde{\Delta}_{\circlearrowright}^{\prime 2} + \kappa^2} n_m.
\label{14}
\end{equation}
Under strong-driving conditions, the steady-state amplitude is given by $|\alpha_{\circlearrowright}| = \sqrt{N}$. Combining Eq.~(\ref{14}), the dimensionless parameter $\eta$ in Eq.~(\ref{11}) is calculated as $\eta \approx 5.53 \times 10^{-5} \ll 1$ taking parameters in Fig.~\ref{3xin.eps}. Consequently, the quantum fluctuations are negligible relative to the steady-state amplitude due to the very small $\eta$, justifying that first-order linearization remains valid.

\end{document}